\documentclass[useAMS,usenatbib,usegraphicx]{mn2e}

\title[Pluto's moon system: survey of the phase space I.]{Pluto's moon system: survey of the phase space I.}
\author[I. Nagy, \'A. S\"uli, B. \'Erdi]{I. Nagy,\thanks{E-mail:
I.Nagy@astro.elte.hu} \'A. S\"uli and B. \'Erdi\\
Department of Astronomy, E\"otv\"os  University, Budapest, Hungary}

\begin{document}

\date{}

\pagerange{\pageref{firstpage}--\pageref{lastpage}} \pubyear{2006}

\maketitle

\label{firstpage}

\begin{abstract}
The dynamical structure of the phase space of the Pluto--Charon system is studied in the model of the spatial circular restricted three-body problem by using numerical methods. With the newly discovered two small satellites S/2005 P1 and S/2005 P2, the Pluto--Charon system can be considered as the first known binary system in which celestial bodies move in P-type orbits. It is shown that the two satellites are in the stable region of the phase space and their origin by capture is unlikely. Also the large mass parameter allows to regard the satellites as a model of a new class of exoplanets orbiting around stellar binary systems.
\end{abstract}


\section{Introduction}

In 1930 C. Tombaugh discovered Pluto, the ninth planet of the Solar system. 
Pluto's first moon, Charon was found by \cite{Christy1978}. The
Pluto--Charon system is remarkable, since in the Solar system Charon is the largest moon relative to its primary, with the highest mass-ratio 0.130137.  Subsequent searches for other satellites around Pluto had been unsuccessful until mid May 2005, when two small satellites were discovered \citep{Stern2005} provisionally designated as S/2005 P1 and S/2005 P2 (henceforth P1 and P2). With this observation Pluto became the first Kuiper Belt object known to have multiple satellites. These new satellites are much smaller than Charon, with  diameters 61-167 km (P1) and 46-137 km (P2) depending on the albedo. Both  satellites appear to be moving in nearly circular orbits in the same orbital plane as Charon, with orbital periods 38 days (P1) and 25 days (P2).

From a dynamical point of view the Pluto--Charon system corresponds to such
a binary system whose mass parameter is approximately one tenth. The phase space of binaries and the Pluto--Charon system can be studied simultaneously. To survey the phase space of binaries is a fundamental task, since more and more exoplanetary systems are being discovered. The great majority of exoplanets have been observed around single stars, but more than 15 planets are already known to orbit one of the stellar components in binary systems (this type of motion is referred to as satellite or S-type motion). Other observational facts also favour the study of the phase space of binaries. Observations show that 60\% of the main sequence stars are in binary or multiple systems \citep{Duquennoy1991}. These facts give grounds for the investigation of the stability properties of planetary orbits in binaries.

With the increasing number of extrasolar planetary systems around single stars,
general stability studies of such systems have become of high concern. Many investigations were conducted for individual systems, especially for
those which harbour multiple planets like {\tt Ups Andromedae} 
see e.g. \citep{Lissauer1999,Laughlin2001,Stepinski2000},
{\tt Gliese 876} see e.g. \citep{Kinoshita2001,Hadjidemetriou2002}
or {\tt HD82943} see e.g. \cite{Hadjidemetriou2002}. 
There are also studies on planetary orbits in binaries. The so far discovered planets in binaries move in S-type orbits. Theoretically there is another possible type of motion, the so-called planetary or P-type, where a planet moves around both stars. There are several studies on S- and P-type motions using the model of the planar elliptic restricted three-body problem see e.g. \citep{Dvorak1984,Dvorak1986,Holman1999,Lohinger2002} and references therein. The three-dimensional case, that is the effect of the inclination was studied by
\cite{Lohinger2003} for P-type orbits in equal-mass binary models.

The main goal of this paper is to investigate the dynamical structure of the phase space of the Pluto--Charon system which can be considered as the first known binary system in which celestial bodies, namely P1 and P2 move in P-type orbits. In Section 2 we describe the investigated model and give the initial conditions used in the integrations. The applied numerical methods are briefly explained in Section 3. The results are shown in Section 4. Section 5 is devoted to some conclusions.

\section{Model and initial conditions}

To study the structure of the phase space of the Pluto--Charon system we applied the model of the spatial circular restricted three-body problem. We integrated the equations of motion by using a Bulirsch--Stoer integrator with adaptive stepsize control. The orbits of the primaries were considered circular and their mutual distance $A$ was taken as unit distance. The orbital plane of the primaries was used as reference plane, in which the line connecting the primaries at $t=0$ defines a reference $x$-axis. We assume that the line of nodes of the orbital plane of the massless test particle (i.e. P1 or P2) coincides with the $x$-axis at $t=0$, thus the ascending node $\Omega=0^{\circ}$. The pericenter of the test particle's orbit is also assumed to be on the $x$-axis at $t=0$, thus the argument of the pericenter $\omega=0^{\circ}$. Though P1 and P2 are in the orbital plane of Charon, still we study the problem more generally by considering the effect of non-zero inclinations on the orbital stability. Thus our results are applicable to a wider class of satellite or planetary systems around binaries for the  mass parameter $\mu=m_2/(m_1+m_2)=0.130137$, corresponding to the Pluto--Charon system ($m_1$ and $m_2$ being the mass of Pluto and Charon, respectively).

To examine the phase space and the stability properties of P-type orbits, 
we varied the initial orbital elements of the test particle in the following way (see Table \ref{tab:1}):
\begin{itemize}
\item the semimajor axis $a$ is measured from the barycentre of Pluto and Charon and it is varied from 0.55 to 5 $A$ (to 4 $A$ in Fig. \ref{fig:1}) with stepsize $\Delta a=0.005$ $A$,
\item the eccentricity $e$ is varied from 0 to 0.3 with stepsize $\Delta e = 0.05$ ($\Delta e = 0.002$ in Fig. \ref{fig:3}),
\item the inclination $i$ is varied from $0^{\circ}$ to $180^{\circ}$ with stepsize $\Delta i=1^{\circ}$,
\item the mean anomaly $M$ is given the values: $0^{\circ}$, $45^{\circ}$, $90^{\circ}$, $135^{\circ}$, and $180^{\circ}$.
\end{itemize}

The above orbital elements refer to a barycentric reference frame, where the mass of the barycentre is $m_1+m_2$. By the usual procedure we calculated the barycentric coordinates and velocities of the test particle and then transformed them to a reference frame with Pluto in the origin. In the numerical integrations we used the latter coordinates and velocities.

\begin{table*}
\centering
\begin{minipage}{140mm}
\caption[]{In the first three rows the orbital elements from unrestricted fits
(epoch = 2452600.5) are listed \citep{Buie2005}: $a$, $e$, $i$, $\omega$, $\Omega$ and $M$ denote the semimajor axis, eccentricity, inclination, argument of the pericenter, longitude of the ascending node, and mean anomaly.
In the last column the orbital periods are given in days. The orbital elements for P- and S-type orbits are given with the corresponding stepsizes.}
\center
\begin{tabular}{l|rrrrrrr}
\hline
Object    & $a$ [$A$] & $e$    & $i [^{\circ}]$       & $\omega [^{\circ}]$ & $\Omega [^{\circ}]$  & $M [^{\circ}]$    & T  [day] \\
\hline
Charon    &  1.0   &  0      &  96.145   &  --      &  223.046  &  257.946 & 6.387 \\
S/2005 P2 &  2.487 &  0.0023 &  96.18    &  352.86  &  223.14   &  267.14  & 25 \\
S/2005 P1 &  3.31  &  0.0052 &  96.36    &  336.827 &  223.173  &  122.71  & 38 \\
\hline
P-type   & 0.55--5 & 0--0.3  & 0--180    &  0       &  0        & 0--180   & \\
$\Delta$  & 0.005  & 0.05    & 1         &  --      &  --       & 45       & \\
\hline
S-type   &0.1--0.9 & 0--0.5  & 0         &   0      &   0       &  0       & \\
$\Delta$  &0.008   & 0.005   & --        &  --      &  --       & --       & \\
\hline
\end{tabular}
\label{tab:1}
\end{minipage}
\end{table*}

In addition to P-type orbits, S-type orbits were also integrated to explore the phase space between Pluto and Charon. The initial conditions for S-type orbits are also listed in Table \ref{tab:1}, these are the same both for orbits around Pluto and around Charon.

In total almost six million orbits were integrated for $10^3$ Charon's period (hereafter $T_C$) and approximately 500 thousand for $10^5$ $T_C$.

\section{Methods}

\begin{enumerate}
\item {\sc The relative Lyapunov indicator (RLI).} To determine the dynamical character
of orbits we used three methods.
The method of the relative Lyapunov indicator (RLI) was introduced by \cite{Sandor2000} for a particular problem, and its efficiency was demonstrated in a later paper \citep{Sandor2004} for 2D and 4D symplectic mappings and for Hamiltonian systems. This method is extremely fast to determine the ordered or chaotic nature of orbits.

The method is based on the idea that two initially nearby orbits are integrated
simultaneously and the evolution of their tangent vectors are also followed.
For both orbits the Lyapunov characteristic indicator (LCI) is calculated and
the absolute value of their difference over time is defined as the RLI:
\begin{eqnarray}
\mathtt{RLI}(t) =\frac{1}{t} |\mathtt{LCI}(x_0) - \mathtt{LCI}(x_0+\Delta x) |,  \label{Eq:1}
\end{eqnarray}
where $x_0$ is the initial condition of the orbit and $\Delta x$ is the distance of the nearby orbit in the phase space. The value of the RLI is characteristically several orders of magnitude smaller for regular than for chaotic orbits, thus they can be distinguished easily.
\item {\sc The maximum eccentricity method (MEM).} For an indication of stability a straightforward check based on the
eccentricity was used. This action-like variable shows the probability of
orbital crossing and close encounter of two planets, and therefore its value
provides information on the stability of orbits. We examined the behaviour of
the eccentricity of the orbit of the test particle along the integration, and used its largest value ME as a stability indicator; in the following we call it the maximum eccentricity method (hereafter MEM). This is a reliable indicator of chaos, since the overlap of two or more resonances induce chaos and large excursions in the eccentricity. We know from experience that instability comes from a chaotic growth of the eccentricity. This simple check was already used in
several stability investigations, and was found to be a powerful indicator of the stability character of orbits \citep{Dvorak2003,Suli2005}.
\item {\sc The maximum difference of the eccentricities method (MDEM).}
We developed this new method and applied it for the first time in this investigation.
Two initially nearby trajectories emanating from a chaotic domain of the phase space will diverge according to the strength of chaos. The divergence manifests itself in the differences between the eccentricities of the orbits and in the angle variables. The more chaotic the system is, the faster the difference in the eccentricities grows. This difference is sensible to the variations around the running average of the eccentricity and depends also on the position along the orbit. Thus if the positions along the two orbits change chaotically, the eccentricities of the two orbits also behave differently and their momentary differences can be large even if the average value of the eccentricity of each orbit remains small. This method characterises the stability in the phase space, whereas the MEM does it in the space of orbital elements. We define the stability indicator MDE as:
\begin{eqnarray}
\mathtt{MDE}(t) = \mathtt{max} |e(t,x_0) - e(t,x_0+\Delta x) |,
\end{eqnarray}
where $x_0$ is the initial condition of the orbit and $\Delta x$ is the distance of the nearby orbit in the phase space. The method of the MDE has the advantage with respect to the MEM that in the case of chaotic orbits the MDE grows more rapidly than the ME, and while the difference between the ME for regular and chaotic motions is only 1-2 orders of magnitude, this can be 4-7 orders of magnitude for the MDE and therefore can be detected more easily.
\end{enumerate}

\section{Results}

\begin{figure}
\includegraphics[width=0.9\linewidth]{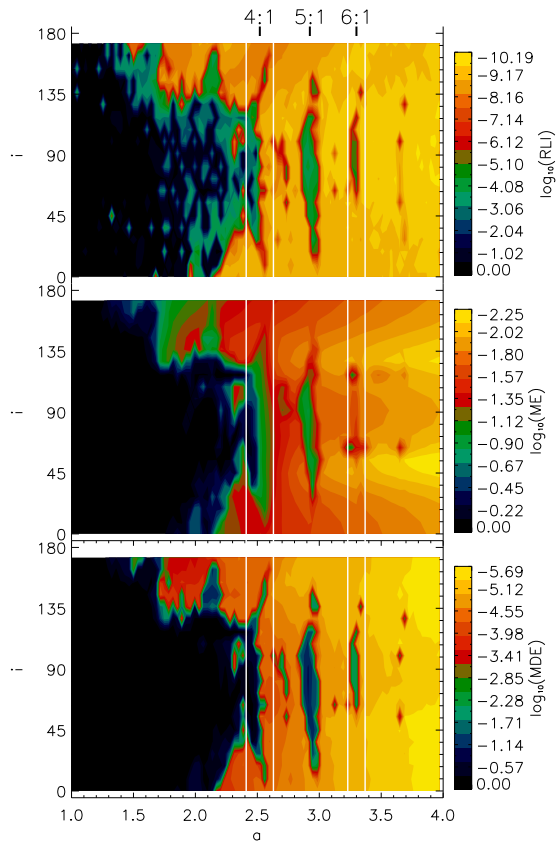}
\caption{The results of the $10^5 T_C$ simulations for $e=0$ in the $a,i$ plane. Black marks chaotic, light grey means stable motion. The white vertical lines indicate the dynamically possible minimal pericenter and maximal apocenter distance of P2 (2.41, 2.63 $A$) and P1 (3.23, 3.37 $A$), see Section 4.2 for details. Above the top panel the main mean motion resonances are marked.}
\label{fig:1}
\end{figure}

We show the results of our investigations in Figs. \ref{fig:1} -- \ref{fig:3}. These were obtained as follows. By varying the initial orbital elements as described in Section 2, we performed the integration of each orbit for five different initial values of the mean anomaly:
$M=0^{\circ},\,45^{\circ},\,90^{\circ},\,135^{\circ}$, and $180^{\circ}$.  For each $M$ the indicators $I^{(M)}(a,e,i)$ were determined, where $I^{(M)}$ stands for RLI, ME, and MDE, respectively. Any value, plotted in the figures, is an average over $M$:
\begin{eqnarray}
\bar{I}(a,e,i) =\frac{1}{5}\sum_{M} {I^{(M)}(a,e,i)}.
\end{eqnarray}
We note, that this averaging in the case of the RLI and the MDE emphasises the chaotic behaviour of an orbit, while in the case of the ME is not so drastic. This is due to the fact, that the difference in the RLI or in the MDE  for regular and chaotic orbits is several orders of magnitude, while in the case of the ME it is only one or two orders of magnitude.

The three methods are not equivalent, however they complete each other. For example, the ME of the Earth is small, indicating stability, although we know from numerical experiments that in fact the Earth is moving on a chaotic trajectory with a small but nonzero Lyapunov exponent. Therefore the ME detects macroscopic instability (which may even result in an escape from the system), whereas the RLI and the MDE are capable to indicate microscopic instability.

In most of the simulations the $I^{(M)}$ values were calculated for $10^3$ $T_C$. To decide whether this time interval is enough to map the real structure of the phase space, several test runs were done for a much longer time span, for $10^5$ $T_C$. 

Fig. \ref{fig:1} summarises the results of the simulations for $10^5$ $T_C$ in the $a,i$ plane, where one can see the structure of the phase space: black for chaotic and light grey for ordered motion. In these simulations $e=0$ was taken for the initial eccentricity of the test particle's orbits.

In general, all the three methods provide the same global structure, however, there are also some details that do not quite coincide. These differences are only natural, since the methods are based on different quantities. In the case of the RLI (top panel) the boundary of the chaotic region is not so sharp as in the cases of the ME (middle) and the MDE (bottom panel). In the RLI map we can see a mixed region (grey colour) in the middle, where chaos is already present, but manifests itself on a longer time-scale (it would need longer integration time to be detected).

The boundary of the chaotic region slightly extends with the increase of
the inclination, starting from 2.15 $A$ at $i=0^{\circ}$. This is in close agreement with the result of \cite{Lohinger2003}, obtained for $\mu=0.5$ (see Fig \ref{fig:1}. of that paper). It reaches its largest extension at
$i \approx 80^\circ$, where it merges with the chaotic region of the 4:1 resonance. Increasing the inclination further, the size of the chaotic region
slowly shrinks. At $i \approx 135^\circ$ the extension of the chaotic zone suddenly drops down to $a = 1.65$ $A$, then it remains nearly constant and its border runs parallel with the $i$-axis. Inspecting Fig. \ref{fig:1}, the above described structure is similar in all panels. 

Several islands appear on the panels, which are connected with mean motion resonances: 4:1 at $a_r=2.52$ $A$, 5:1 at $a_r=2.92$ $A$, and 6:1 at $a_r=3.3$ $A$. The first island is the most significant and as mentioned above it is connected to the large chaotic sea. It is well visible from the figure that the higher the order of the resonance, the smaller is the size of the corresponding island. We note that the width of the islands is larger when the indicators $I^{(M)}$ are averaged over $M$, than for individual values of $M$. The widening effect is a consequence of a shift of the location of the resonance $a_r$, since $a_r$ slightly depends on $M$. The location of a resonance depends also on the inclination. In the case of the 4:1 resonance $a_r$ decreases as $i$ increases, reaching its minimum value at about 80$^\circ$. Further increasing the inclination, $a_r$ begins to grow and reaches its initial value at $i \approx 160^\circ$. This dependence of $a_r$ on $i$ is only less visible in the case of the 5:1 and 6:1 resonances.

Fig. \ref{fig:1} was obtained from simulations for a time span of $10^5$ $T_C$.
Comparing Fig. \ref{fig:1} with the left panel of Fig. \ref{fig:2} ($e=0$), obtained from simulations for $10^3$ $T_C$, a close agreement can be observed. The structures described in Fig. \ref{fig:1} are already visible in Fig. \ref{fig:2} for the shorter time span. Thus we can conclude that the phase space of the Pluto--Charon system can be surveyed in a reliable way by using a time span of $10^3$ $T_C$.

\subsection{The phase space of the Pluto--Charon system}

\begin{figure*}
\centering
\includegraphics[width=1\linewidth]{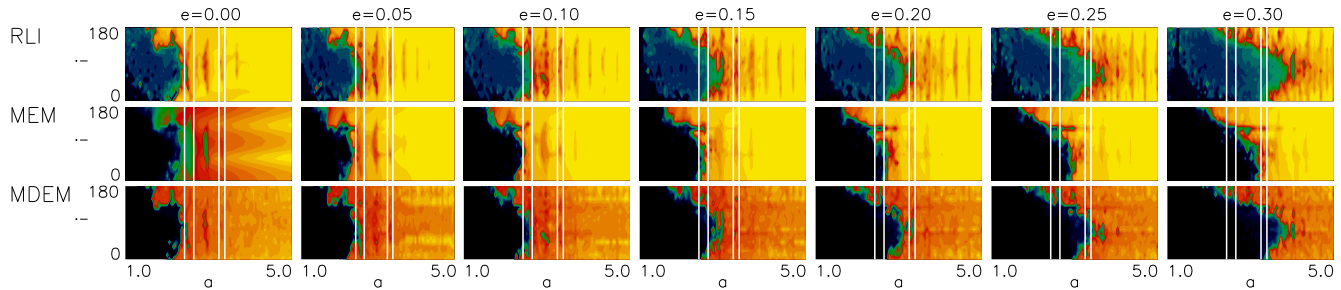}
\caption{The results of the $10^3$ $T_C$ simulations for $e=0,\ldots,0.3$ in the $a,i$ plane. See text for details.}
\label{fig:2}
\end{figure*}

We investigated the behaviour of P-type orbits systematically by changing the initial orbital elements of the test particle as described in Section 2 (see also Table \ref{tab:1}). Beside direct orbits ($i<90^{\circ}$) we studied also
retrograd P-type motion ($i>90^{\circ}$) of the test particle. All the integrations were made for $10^3$ $T_C$. The results are shown in Fig. \ref{fig:2}, where the indicators $\bar{I}$ are plotted on the $a,i$ plane for different values of $e$.

In general, the results show an increase of the chaotic area for higher eccentricities: for $i<160^{\circ}$ the chaotic region grows with $e$.
However, the rate of the increase strongly varies with the inclination.
It can also be seen that the resonant islands merge with the growing
chaotic zone. The most striking feature is that the stability of the retrograd P-type motion practically does not depend on $e$. Inspecting Fig. \ref{fig:2} it is evident that the border of the chaotic zone for $i> 160^{\circ}$ stay almost constantly at $a \approx 1.7$ $A$.

Simulations to investigate the phase space between Pluto and Charon were also performed for time spans $10^4$ $T_C$. We examined S-type orbits around Pluto and Charon (see the initial conditions in Table 1, where $a$ was measured from the respective celestial body). In the case of Pluto we found that stable orbits can exist up to 0.5 $A$, but not above this limit. No stable S-type orbits were found around Charon.

\subsection{Stability of the satellites P1 and P2}

\begin{figure}
\includegraphics[width=8cm]{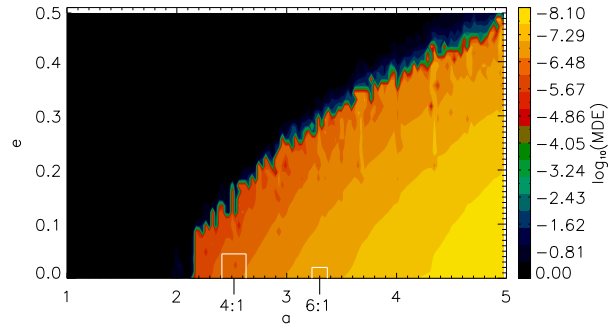}
\caption{Stability map in the $a,e$ plane. The dark area is unstable, the grey regions are stable. }
\label{fig:3}
\end{figure}

We have addressed the problem of stability of the recently discovered satellites of Pluto. The results are shown in Fig. \ref{fig:3}, where the values of the MDE (computed for $10^3$ $T_C$ and averaged over for the mean anomalies) are plotted on the $a,e$ plane for the planar case ($i=0^{\circ}$). Below $a=2.15$ $A$
the system is unstable for all $e$, above $a=2.15$ $A$ there is a stable region depending on $e$. The two satellites are situated here, in the small rectangles, indicating their dynamically possible most probable places of occurance. These rectangles are defined by the ME, computed in the vicinity of each satellite. This means that we took a grid around the present value of $a$ of each satellite
with a stepsize $\Delta a=0.005$ $A$, $\Delta i=1.25^{\circ}$ in the interval $i=0-180^{\circ}$, and with initial $e=0$ we computed the largest ME$_{max}$ during $10^5$ $T_C$ (including averaging over the five values of $M$). We obtained that ME$_{max}=0.045$ for P2 and ME$_{max}=0.02$ for P1. In Fig. \ref{fig:3} these values give the height of the rectangles. We computed the possible minimal $r_p=a(1-\mathtt{ME}_{max})$ pericenter and maximal $r_a=a(1+\mathtt{ME}_{max})$ apocenter distances of the satellites, these are 2.41 and 2.63 $A$ for P2 and 3.23 and 3.37 $A$ for P1. These values define the horizontal limits of the rectangles in Fig. \ref{fig:3}, and also the places of the vertical lines in Figs. \ref{fig:1} and \ref{fig:2}.

From Fig. \ref{fig:3} it can be seen that the determined orbital elements of the two satellites are well in the stable domain of the phase space. If P2 and P1 move in the orbital plane of Charon, their eccentricities cannot be larger than 0.17 and 0.31, respectively. The present semimajor axis $a$ of P2 is very close to the 4:1 resonance, whereas that of P1 is close to 6:1. The locations of the exact resonances are well inside the small rectangles. We presume that these satellites probably move in resonant orbits. This could be confirmed by new observations.

\section{Conclusions}

Up to now the Pluto--Charon system is the only known binary system which has a relatively large mass-ratio and celestial bodies revolve around it in P-type orbits. This circumstance and the high ratio of binary stellar systems among stars make it important to study the stability properties of P-type orbits in binaries and particularly in the Pluto--Charon system. Our investigations show that the stable region is wider for retrograd than for direct P-type orbits. With the increase of the eccentricity the chaotic region becomes larger, and because of it the eccentricities of the two satellites, at their present semi-major axis, cannot be higher than 0.17 for P2 and 0.31 for P1. Below $a=2.15$ $A$ orbits are unstable for all eccentricities, thus no satellite could exist here.

\cite{Stern2005} has shown that P1 and P2 were very likely formed 
together with Charon, due to a collision of a large body with Pluto, from material ejected from Pluto and/or the Charon progenitor. This is based on the facts that P1 and P2 move close to Pluto and Charon in nearly circular orbits in the same orbital plane as Charon, and they are also in or close to higher-order mean motion resonances.

Our results are also against the capture origin of these satellites. Firstly, since the stability region for retrograd orbits is wider, it would have been more probable for the satellites to be captured into retrograd than for direct orbits. Secondly, capture into orbits close to the Pluto--Charon binary
cannot be with high eccentricity ($e>0.17$ and 0.31 at the present semimajor axis of P2 and P1), since these orbits become unstable on a timescale of $10^3$ $T_C$. On the other hand, for eccentric capture orbits the tidal circularisation time for P1 and P2 is much longer than the age of the Solar System as \cite{Stern2005} pointed out.

Taking that P1 and P2 were formed together with Charon it may be assumed that other satellites were also formed from the collision material. However, as Fig. \ref{fig:2} shows, to $a \approx 2$ $A$ binary separation for all inclinations and eccentricities the orbits become unstable in a short timescale due to close  encounters with Pluto or Charon. Thus most of the ejected material must have been accumulated onto Pluto or Charon. How P1 and P2 evolved to their current orbits is an unsolved problem. 

\cite{Stern2005} also suggested that the stochastic bombardment of P1 and P2 by small Kuiper Belt debris can generate transient, dusty ice particle rings around Pluto between the orbits of P1 and P2. If these particles are co-planar with the satellites, their eccentricities must be below 0.17-0.31 depending on their semimajor axis as Fig. \ref{fig:3} shows, otherwise they would become unstable in $10^3$ $T_C$.

The existence of S/2005 P1 and S/2005 P2 shows that there can be a new class of exoplanets which revolve in P-type orbits in binary stellar systems. These planetary systems can be more similar to our Solar System as compared to the known exoplanetary systems, since there cannot be large eccentricity orbits in the inner part of these systems.

\section*{Acknowledgments}

Most of the numerical integrations were accomplished on the NIIDP (National Information Infrastructure Development Program) cluster grid system in Hungary.
The support of the Hungarian NRF, grant no. OTKA T043739 is also acknowledged.

\end{document}